\documentclass[]{aastex631}

\graphicspath{{./}{pics/}}
\begin{document}

\title{Sequential Remote Brightenings and Co-spatial Fast Downflows during Two Successive Flares}

\author{B. T. Wang}
\affiliation{School of Astronomy and Space Science, Nanjing University, Nanjing 210023, People's Republic of China; xincheng@nju.edu.cn}
\affiliation{Key Laboratory of Modern Astronomy and Astrophysics (Nanjing University), Ministry of Education, Nanjing 210093, People's Republic of China}
\author{X. Cheng}
\affiliation{School of Astronomy and Space Science, Nanjing University, Nanjing 210023, People's Republic of China; xincheng@nju.edu.cn}
\affiliation{Key Laboratory of Modern Astronomy and Astrophysics (Nanjing University), Ministry of Education, Nanjing 210093, People's Republic of China}
\author{C. Li}
\affiliation{School of Astronomy and Space Science, Nanjing University, Nanjing 210023, People's Republic of China; xincheng@nju.edu.cn}
\affiliation{Key Laboratory of Modern Astronomy and Astrophysics (Nanjing University), Ministry of Education, Nanjing 210093, People's Republic of China}
\author{J. Chen}
\affiliation{School of Astronomy and Space Science, Nanjing University, Nanjing 210023, People's Republic of China; xincheng@nju.edu.cn}
\affiliation{Key Laboratory of Modern Astronomy and Astrophysics (Nanjing University), Ministry of Education, Nanjing 210093, People's Republic of China}
\author{M. D. Ding}
\affiliation{School of Astronomy and Space Science, Nanjing University, Nanjing 210023, People's Republic of China; xincheng@nju.edu.cn}
\affiliation{Key Laboratory of Modern Astronomy and Astrophysics (Nanjing University), Ministry of Education, Nanjing 210093, People's Republic of China}

\begin{abstract}

Remote brightenings often appear at outskirts of source active regions of solar eruptive events, nevertheless, their origin remains to be ascertained. 
In this study, we report imaging and spectroscopic observations of sequential remote brightenings with a combination of H$\alpha$ Imaging Spectrograph (HIS) onboard the Chinese H$\alpha$ Solar Explorer (CHASE), which is the first space-based solar telescope of China, and the Solar Dynamics Observatory. It is found that, during two successive M-class flares occurring on 2022 August 17, multiple ribbon-like brightenings appeared in sequence away from the flaring active region. Meanwhile, abundant cool filament materials moved downward to the sequential remote brightenings as visible at the H$\alpha$ red wing with a line-of-sight speed up to 70 km s$^{-1}$.
The extrapolated three-dimensional magnetic field configuration shows that the sequential remote brightenings correspond to the footpoints of closed ambient field lines whose conjugate footpoints are rooted in the main flare site. We suggest that the sequential remote brightenings are most likely caused by the heating of interchange reconnection between the erupting flux rope and closed ambient field, during which the rope-hosting filament materials are transferred to the periphery of flaring active region along the closed ambient field rather than to the interplanetary space like in the scenario of the slow solar wind formation.

\end{abstract}


\section{Introduction} \label{sec:intro}
Solar flares are phenomena of explosive magnetic energy releases that frequently occur in solar atmosphere and produce electromagnetic emissions at a wide range of wavelengths \citep[for reviews see][]{Fletcher2011,Benz2018}. Flares are usually classified into two types, i.e., confined and eruptive flares, the latter of which are accompanied with coronal mass ejections \citep[CMEs;][]{Schmieder2015}. In the standard model \citep{Carmichael1964,Sturrock1966,Hirayama1974,Kopp1976}, flares and CMEs are considered to be two manifestations of the same process \citep{Zhang2001}, namely the violent disruption of coronal magnetic field. Magnetic reconnection taking place in the wake of erupting CMEs is believed to play a crucial role in converting magnetic energy into kinetic energy of CMEs, thermal energy of flares and energetic particles \citep{Priest2002,Shibata2011}. In observations, the most remarkable characteristic of flares are the brightening ribbons at multi-passbands such as H$\alpha$ and ultraviolet (UV), corresponding to the footpoints of post-flare loops heated by energetic particles \citep{Benz2018}. It is believed that the particles propagate along reconnection-formed field lines to the lower atmosphere and heat the plasma over there through Coulomb collisions \citep[e.g.,][]{Tian2018}.

Aside from the magnetic reconnection in the current sheet connecting CMEs and flare loops \citep{Cheng2018}, it may occur between the erupting CMEs and ambient/background field, namely the interchange reconnection \citep{Crooker2002}, which was proposed originally to interpret the origin of slow solar wind \citep[e.g.,][]{Bale2019}. It was also suggested that interchange reconnection may occur between the CME magnetic flux rope (MFR) and the open field during the eruption so as to facilitate the escape of energetic particles \citep[e.g.,][]{Masson2013, Yang2015}. Two pieces of indirect evidence are observations of type III radio bursts caused by escaped electron beams, in particular for those extending to the wavelength of decamneter-kilometer \citep{van2008,Reid2014,kou2020,Wang2022}, and intense solar energetic particles (SEPs) that arrive at the Earth and cause space weather effects \citep{Masson2012,Masson2013}. Moreover, interchange reconnection is also able to produce the leakage of filament materials once the eruption involves filaments, as well as the drifting of filament footpoints \citep{Aulanier2019,Yan2020,Li2022}.
In addition, interchange reconnection was also proposed to interpret the evolution of the boundary of coronal holes \citep{Subramanian2010,Moore2015} and the formation of switchbacks in solar wind \citep{Drake2021}.

In this study, we report imaging and spectroscopic observations of ribbon-like remote brightenings, which propagated in a jumped manner away from associated active region during an eruption. With careful analyses, we suggest that these sequential remote brightenings are most likely to be caused by the interchange reconnection between erupting MFR and the closed ambient field. The data and methods are introduced in Section \ref{sec:style}. The properties of sequential remote brightenings and related magnetic configuration are shown in Section \ref{sec:floats}, which is followed by a summary and discussion in Section \ref{sec:cite}.

\section{Instruments and Methods} \label{sec:style}
The data applied in this study is mainly acquired from the H$\alpha$ Imaging Spectrograph (HIS), which is equipped on the Chinese H$\alpha$ Solar Explorer \citep[CHASE;][]{LiC2022}. The CHASE/HIS provides full-disk scanning of the Sun at wavebands of H$\alpha$ (6559.7 - 6565.9 \AA) and Fe I (6567.8 - 6570.6 \AA) with a spectral sampling of 0.024 \AA. A full-Sun scanning takes $\sim$ 46 s and the time interval of adjacent scanning, equivalent to temporal cadence, is $\sim$ 73 s. In the binning mode data applied in present study, the spatial resolution is $\sim$ $2.0^{\prime \prime}$ and the spectral sampling is 0.048 \AA. The Atmospheric Imaging Assembly \citep[AIA;][]{Lemen2012} on board the Solar Dynamics Observatory (SDO) provides UV and EUV images of the solar atmosphere at nine passbands. The temporal cadence is 12 s (94, 131, 171, 193, 211, 304 and 335 \AA) or 24 s (1600 \AA\ and 1700 \AA) and the spatial resolution is $1.5^{\prime \prime}$. The Helioseismic and Magnetic Imager \citep[HMI;][]{Scherrer2012} provides the full-disk line-of-sight (LOS) magnetograms with the spatial resolution of $1^{\prime \prime}$ but a lower time cadence ($\sim$ 45 s) than that for AIA. The solar full-disk H$\alpha$ images from the Global Oscillation Network Group \citep[GONG;][]{Harvey1996} and the soft X-ray (SXR) images from the X-Ray Telescope (XRT) onboard Hinode \citep{Golub2007} are also used.

We first calibrate CHASE/HIS data, which includes the dark-field and flat-field correction, slit image curvature correction, wavelength and intensity calibration, and coordinate transformation \citep{Qiuy2022}. To make alignment of the HIS images with the SDO/AIA images, we use the routine \textit{SIR.pro} \citep{Fengs2012,Yangyf2015}, which is based on the cross correlation method. The alignment is achieved by comparing the HMI continuum images and HIS Fe I line image. Moreover, we identify the sequential remote brightenings using the HIS H$\alpha$ line center images. With a back and forth test, we find that the regions whose intensity is  30\% larger than the full-disk average intensity can well define the sequential remote brightenings. The EUV images observed by SDO/AIA are enhanced using the multi-scale Gaussian normalization technique \citep{Morgan2014}. For UV passbands, we also use the 1600 \AA\ to 1700 \AA\ ratio images to enhance the visibility of sequential remote brightenings at 1600 \AA.

\begin{figure}[ht!]

{\centering\includegraphics[width=1.0\textwidth]{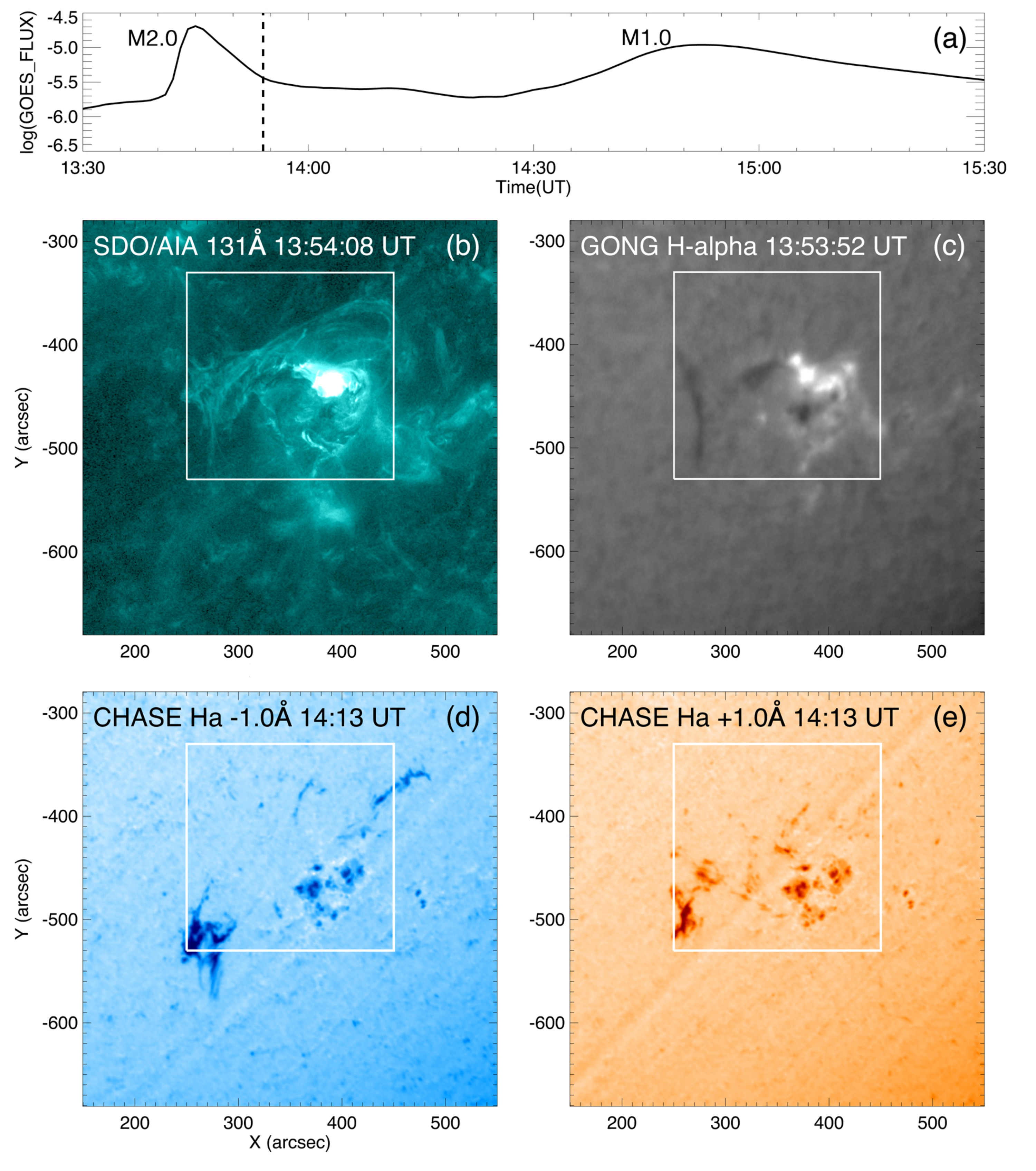}

}
\caption{(a) The GOES 1-8 \AA\ SXR light curve showing two M-class flares occurred successively during this eruptive event. (b)-(c) SDO/AIA 131 \AA\ and GONG H$\alpha$ images at 13:54 UT (the vertical dashed line in panel a) showing the hot-channel-like MFR and the eruptive filament materials, respectively. (d)-(e) CHASE H$\alpha$ blue wing (-1 \AA) and red wing (+1 \AA) images showing the upward- and downward-moving filament materials, respectively. The white boxes mark the region containing AR 13078 and remote brightenings. An animation of panels a-c is available to show the temporal evolution of the eruption. The duration of the animation is 6 s. The animation begins at 13:32 UT and ends at 15:30 UT. 
\label{fig:7}}
\end{figure}

\begin{figure}[ht!]

{\centering\includegraphics[width=1.0\textwidth]{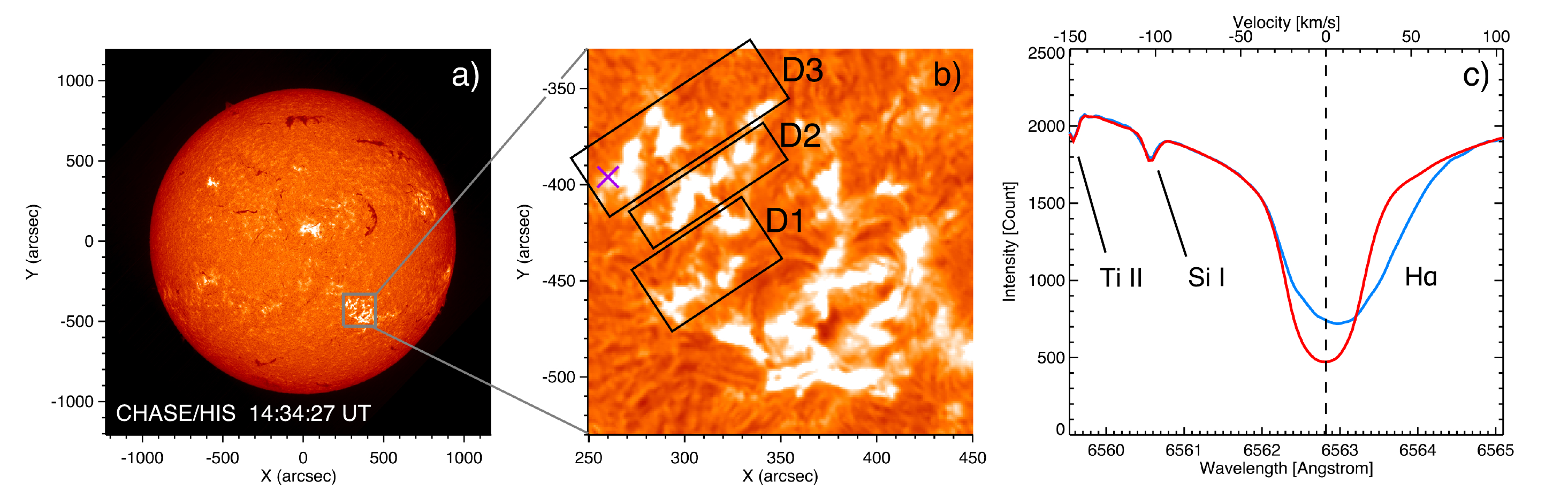}

}
\caption{(a) Solar chromosphere observed at the H$\alpha$ line center during the scanning at 14:34:27 - 14:35:13 UT on 2022 August 17. (b) Zoom-in region of interest (the grey box in panel a) showing the sequential remote brightenings nearby the main flare site at the south-west on the Sun. The black boxes D1, D2 and D3 address the sequential ribbon-like remote brightenings. (c) The H$\alpha$ line profile (including Ti II and Si I lines) at one pixel in the sequential remote brightenings (marked as the purple cross in panel b) and the profile averaged over the quiescent region. 
\label{fig:1}}
\end{figure}

\begin{figure}[ht!]

{\centering\includegraphics[width=1.0\textwidth]{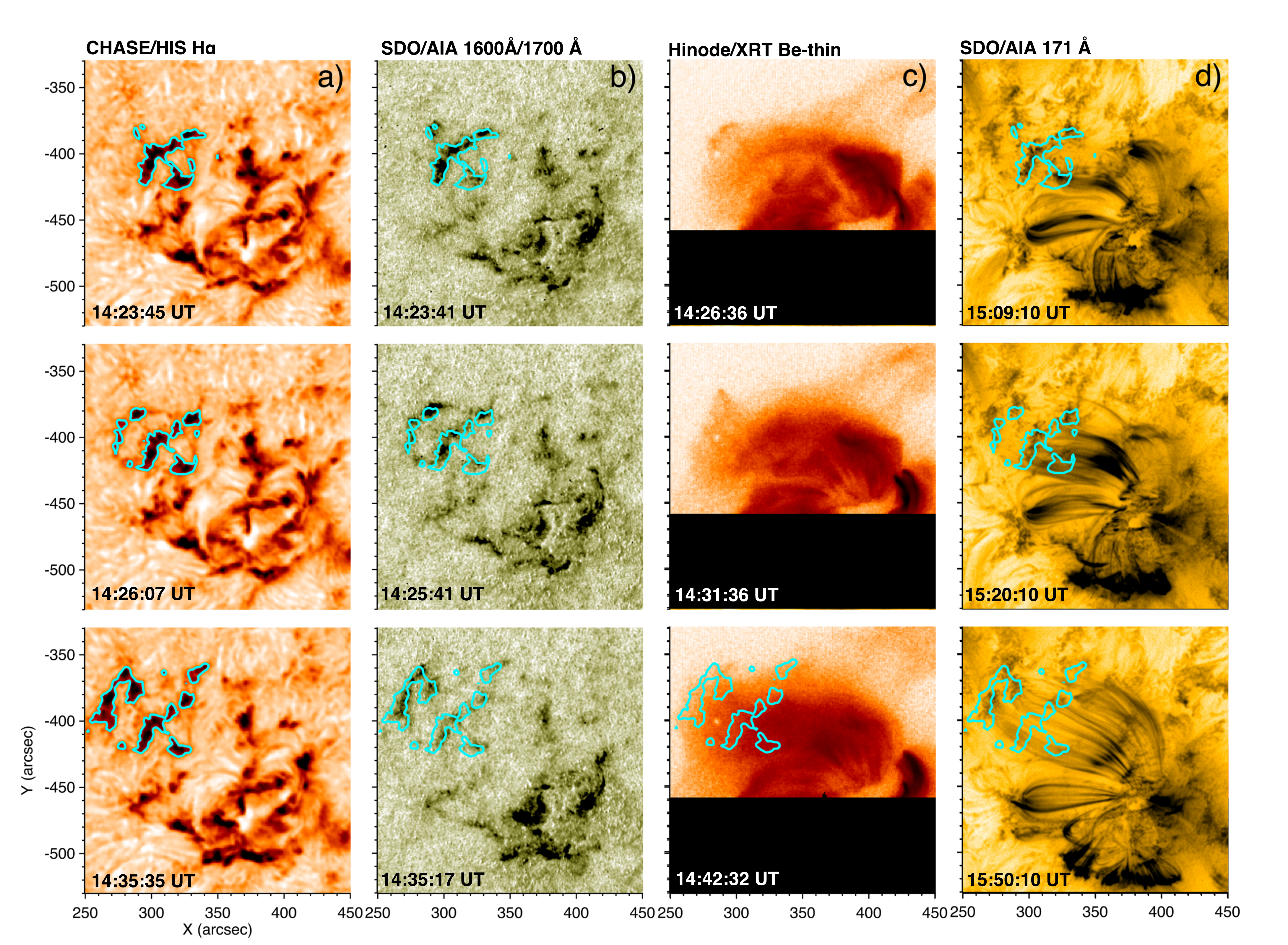}

}
\caption{(a)-(b) Reverse images at the H$\alpha$ line center and ratio images of 1600 \AA\ to 1700 \AA\ showing the temporal evolution of the sequential remote brightenings (white boxes in Figure \ref{fig:7}b-c). (c)-(d) 171 \AA\ and Hinode/XRT Be-thin reversed images showing bright loops connecting the flaring active region to the sequential remote brightenings (white boxes in Figure \ref{fig:7}b-c). Cyan curves mark the boundaries of the remote brightenings.
\label{fig:2}}
\end{figure}

We determine the Doppler shifts of the H$\alpha$ line as follows. First, we get the contrast spectra $C(x, \lambda)$ via
$C(x, \lambda)=\left[I(x, \lambda) - I_{0}(\lambda)\right]/I_{0}(\lambda)$, where $I(x, \lambda)$ is the spectral line intensity at the pixel $x$, $I_{0}(\lambda)$ is the averaged line intensity of the quiescent region with a size of $50^{\prime \prime} \times 50^{\prime \prime}$ near the eruption site. Then we calculate the centroid of the strongest absorption dip at the red wing of $C(x, \lambda)$. Finally, the Doppler velocity is estimated by considering that this red wing absorption is caused by downward moving cool plasma.

\section{Results} \label{sec:floats}
\subsection{Overview of the eruptive event} \label{subsec:tables}

The timeline of the eruptive event is shown in Table \ref{tab:1}. On 2022 August 17, an M2.0 flare occurred in NOAA active region (AR) 13078 with its GOES SXR flux peaking at 13:45 UT (Figure \ref{fig:7}a). Accompanied with the flare, a rising channel-like structure, known as an MFR \citep{Cheng2013,Cheng2017}, was observed at the AIA 131 \AA\ (Figure \ref{fig:7}b). With the rising and expansion of the MFR, a Morton wave appeared at 13:44 UT and then propagated towards the northeast lasting for several minutes. The early eruption of the MFR is found to experience two periods. It first propagated towards the northeast but suddenly changed the direction to the southwest around 13:50 UT. At $\sim$ 14:15 UT, a bright thin thread as part of the MFR can be observed at the northeast of the ARs. At 14:30 UT, the second M1.0 class flare from the same AR started to occur with the flux peaking at 14:52 UT. During the same time, the MFR mainly moved towards the southwest with its left leg clearly visible at almost all EUV passbands. After 15:00 UT, a halo CME was observed by the C2 telescope of the Large Angle and Spectrometric Coronagraph (LASCO) on board Solar and Heliospheric Observatory (SOHO). With a careful inspection, an exceptional feature is that two filament segments were involved in the MFR eruption. The first filament segment from the east of the AR rose along with the left leg of the MFR (Figure \ref{fig:7}c). After 14:10 UT, the second segment from the northwest of the AR rose with the right leg of the MFR. These two filament segments sharply erupted as a whole with the MFR after the eruption changed the propagation direction with a large amount of filament materials moving outward as visible at the H$\alpha$ blue wing image (Figure \ref{fig:7}d).

An interesting phenomenon is that some remote brightenings appeared sequentially at multi-passbands (including X-ray, EUV, UV and H$\alpha$ line center) after the MFR changed its original eruptive orientation ($\sim$ 14:15 UT -- 14:45 UT). Figure \ref{fig:1} presents the chromospheric images of the remote brightenings at the H$\alpha$ line center and their representative H$\alpha$ line profile. It is clear that the remote brightenings appeared as multiple ribbon-like brightenings at the periphery of source active region (Figure \ref{fig:1}b) and the corresponding H$\alpha$ line showed obvious red asymmetry (Figure \ref{fig:1}c). 
We argue that both the remote brightenings and the second flare were caused during the second stage of the MFR eruption. In the paper, we mainly focus on the formation of these interesting remote brightenings.

\begin{deluxetable*}{ll}
\tablenum{1}
\tablecaption{Timeline for the Eruption on 2022 August 17\label{tab:1}}
\tablewidth{0pt}
\tablehead{
\colhead{Time} & \colhead{Observations}
}
\startdata
13:26 UT & The first M2.0 class flare started \\
13:26 - 13:50 UT & The hot-channel-like MFR appeared and rose \\
13:44 UT & A Morton wave formed and propagated towards the northeast \\
13:45 UT & The first flare peaked \\
After 13:50 UT & The MFR changed its propagation direction to the southwest, finally giving rise to a CME \\
14:15 - 14:45 UT & Sequential remote brightenings appeared \\
After 14:26 UT & Sequential bright loops appeared in the X-ray images \\
14:52 UT & The second M1.0 class flare peaked \\
\enddata
\end{deluxetable*}

\begin{figure}[ht!]

{\centering\includegraphics[width=1.0\textwidth]{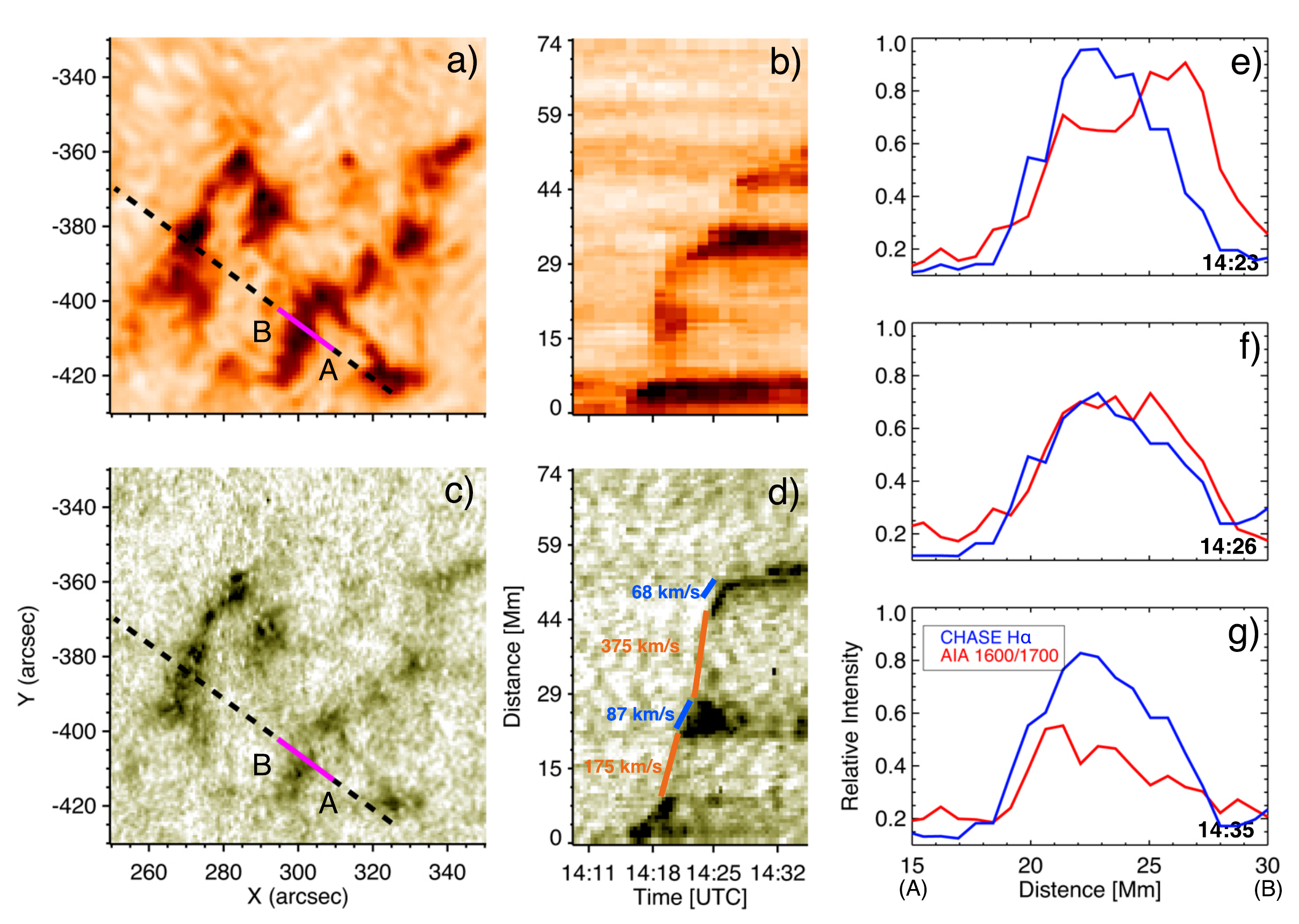}

}
\caption{(a) H$\alpha$ line center image of the sequential remote brightenings at 14:35 UT. (b) Distance-time plot of the slit (the dashed line in panel a) image at the H$\alpha$ line center showing the propagation of remote brightenings along the slit. (c)-(d) same as (a)-(b) but for ratio images of 1600 \AA\ to 1700 \AA. The velocities of the remote brightening front propagating within ribbons (orange) and jumping from one ribbon to another (blue) are denoted in panel d. (e)-(g) Distribution of the H$\alpha$ line center intensity and the 1600 \AA\ to 1700 \AA\ ratio along the segment of the slit denoted in pink in panels a and c.  \label{fig:3}}
\end{figure}

\begin{figure}[ht!]

{\centering\includegraphics[width=1.0\textwidth]{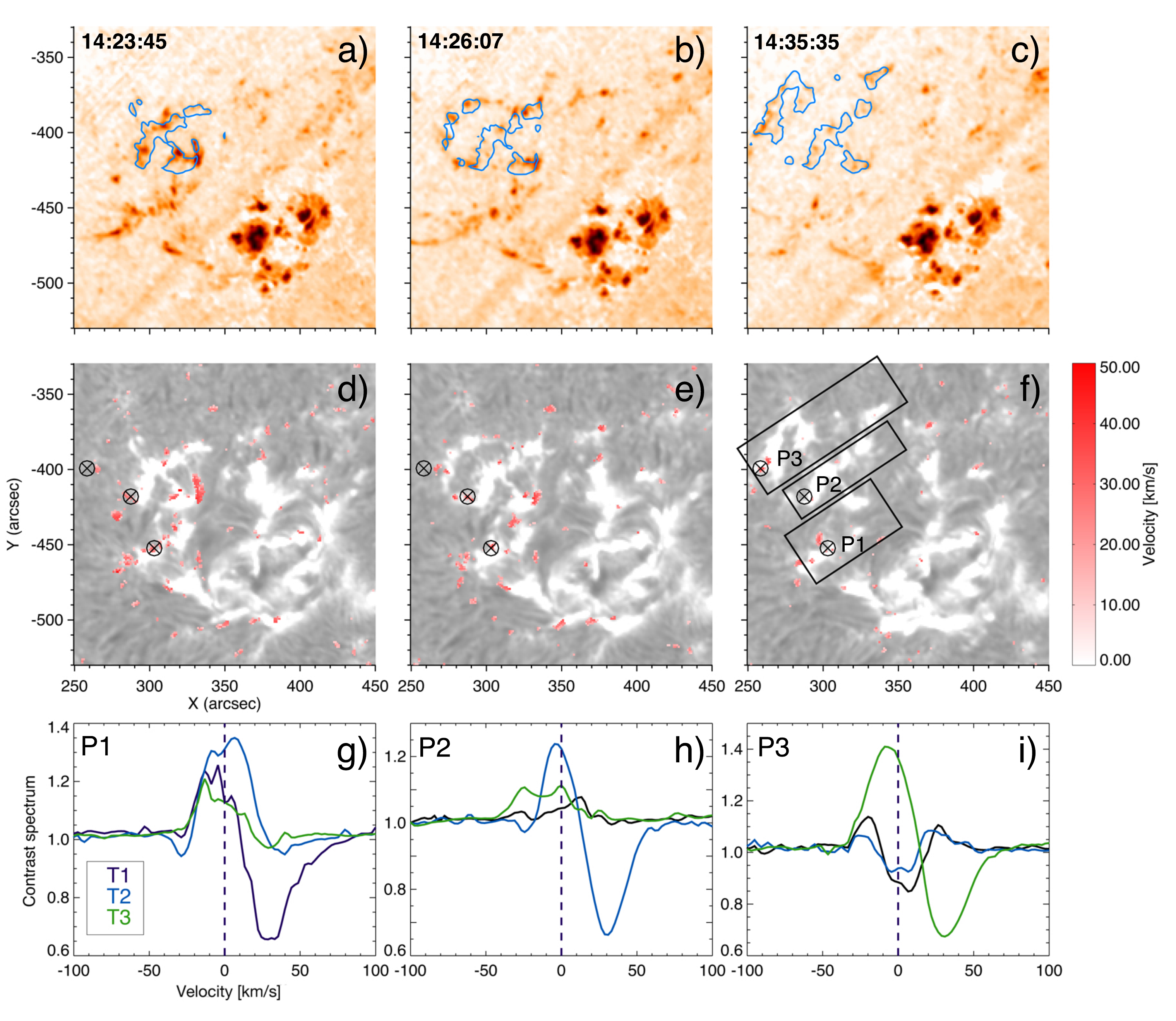}

}
\caption{(a)-(c) Time sequences of H$\alpha$ red wing (+1 \AA) images showing the downward-moving cool materials. The blue curves are the boundaries of sequential remote brightenings. (d)-(e) H$\alpha$ line center images overlaid by patches with significant Doppler velocities caused by downward-moving cool materials. P1, P2 and P3 are three pixels at the sequential remote brightenings. (g)-(i) Contrast profile of H$\alpha$ at P1, P2 and P3 at three moments showing an obvious absorption at the H$\alpha$ red wing. The vertical dashed line represents the H$\alpha$ line center. An animation of panels a-c and d-f is available to show the evolution of downward-moving cool materials. The duration of the animation is 4 s. The animation begins at 14:10 UT and ends at 14:35 UT. 
\label{fig:4}}
\end{figure}

\begin{figure}[ht!]

{\centering\includegraphics[width=1.0\textwidth]{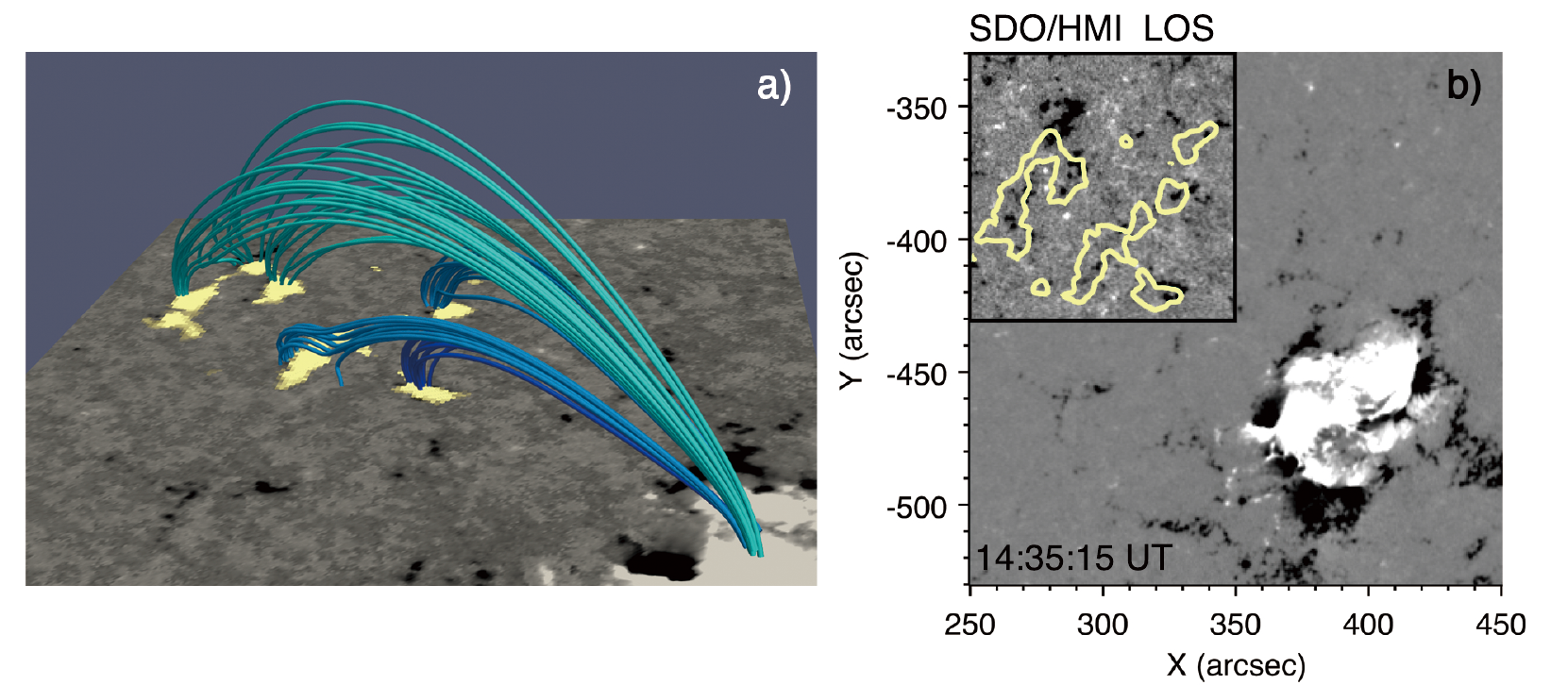}

}
\caption{(a) Extrapolated 3D potential field configuration of the closed ambient magnetic loops that connect the sequential remote brightenings and the AR 13078. The background is the LOS magnetogram overlaid by sequential remote brightenings (yellow). The loops in dark blue, blue, cyan root in remote brightenings D1, D2 and D3 as shown in Figure \ref{fig:1}, respectively. (b) LOS magnetogram of the region we currently study. The maximum unsigned magnetic field strength is 300 G and that in remote brightening regions (marked as black boxes) is 50 G. Yellow curves in the sub-panel mark the boundaries of the remote brightenings.  \label{fig:5}}
\end{figure}

\subsection{Features of the sequential remote brightenings}  \label{sec:A}
We identify three remote brightenings (D1, D2 and D3) according to their distances to the main flare region. The closest D1 started to appear at about 14:15 UT. Then D2 and D3 occurred sequentially within 30 minutes. For each remote brightening, it appeared almost simultaneously at different passbands, including H$\alpha$, UV, EUV and X-ray. However, the remote brightenings disappeared first at the higher-temperature passbands and then the lower ones, and thus their duration times are quite different at different passbands. Specifically, the sequential remote brightenings lasted for 55 minutes at the 335 \AA\ passband, 65 minutes at the 304\AA\ passband, and about 70 minutes at the H$\alpha$ line center. Moreover, after the appearance of sequential remote brightenings, some bright loops connecting the remote brightenings and the flare site started to be visible in the X-ray passband from 14:26 UT and then sequentially from the high- to low-temperature EUV passbands (Figure \ref{fig:2}c-d). These characteristics highly resemble those of the main flare process, during which the bright flare ribbons correspond to the footpoints of post-flare loops that gradually cool down from an initially high temperature to a later low temperature. 

Note that these loops are obviously different from the post-flare loops pertinent to the two M-class flares spatially and temporally. Hence, we argue that magnetic reconnection, which is suspected to be interchange reconnection as interpreted in Section \ref{sec:cite}, was likely to take place between the MFR and the ambient field. Such a reconnection process, being distinguished from the flare reconnection above the post-flare loops, produced sequential remote brightenings.
The H$\alpha$ line center intensity of sequential remote brightenings is 150\% times the average intensity of the quiescent region. This value is significantly smaller than that for the main flare ribbons, suggesting that the reconnection causing remote brightenings is much milder than that causing the main flare. The spatial extensions of remote brightening regions are about $100^{\prime \prime} \times 100^{\prime \prime}$, which is comparable to the size of the main AR. The remote brightenings are mostly ribbon-like brightenings, whose widths are estimated to vary from 5 Mm to 20 Mm. 
In addition, we also estimate the propagation velocity of remote brightenings based on the distance-time plots as shown by panels b and d in Figure \ref{fig:3}. The velocity varies with propagation directions but is mainly in the range of $\sim$50--120 km s$^{-1}$. One important feature for these remote brightenings is that the remote brightening leading front seems to jump between different ribbons with an apparent velocity of $\sim$150--400 km s$^{-1}$, which is obviously larger than the propagation speed of the remote brightening leading front within the ribbons.

To reveal the differences of sequential remote brightenings at the H$\alpha$ and UV 1600 \AA\ bands, we plot the distribution of the intensity along the slit (as shown by the pink segments of the dashed line in Figure \ref{fig:3}a-b). First, we find that the intensity of sequential remote brightenings at the 1600 \AA\ passband decreases faster than that in the H$\alpha$ line center (Figure \ref{fig:3}e-g). Second, it seems that the leading fronts of remote brightenings at the 1600 \AA\ propagated faster than those at the H$\alpha$. However, the distances between the fronts at two passbands are less than 5 Mm for almost all directions (Figure \ref{fig:3}e). When the leading fronts stopped propagation, such discrepancies disappeared quickly.

A more important finding is that the sequential remote brightenings were accompanied with downward-moving filament materials (Figure \ref{fig:4}a-c). In the H$\alpha$ red wing images, one can observe some dark patches moving towards the northeast. They usually lasted for several minutes. Like the propagation of remote brightenings, these patches first appeared at D1 and then at D2 and D3 in sequence. The second important property is that most dark patches coincide with the remote brightenings not only spatially but also temporally. Figure \ref{fig:4}g-i displays the H$\alpha$ contrast profiles of these dark patches, which show significant absorption at the H$\alpha$ red wing. Using the contrast profiles, we derive the Doppler velocities of the absorption dip at the red wing, which are up to 70 km s$^{-1}$ (Figure \ref{fig:4}d-f). This implies cool filament materials moving downward close to the remote brightenings.

\subsection{Magnetic Property of the sequential remote brightenings} \label{sec:M}
The HMI LOS magnetograms disclose that the AR 13078 consists of a main positive polarity region surrounded by several discrete negative polarity regions. The sequential remote brightenings are found to be located at regions containing some tiny negative polarity patches outside of the main AR as shown in Figure \ref{fig:5}b, where the magnetic field strength is about 100 G. 

To explore the cause of sequential remote brightenings, we extrapolate the 3D potential field of the AR utilizing the Green function method and investigate the corresponding magnetic connectivity (Figure \ref{fig:5}a). We find large-scale closed ambient loops with their left footpoints well corresponding to the remote brightenings and the right ones being located close to the main flare region. Moreover, these ambient loops are highly consistent with the bright loops as visible at the AIA EUV passbands (Figure \ref{fig:2}d). We suggest that the interchange reconnection most likely occurs between the erupting flux and large-scale closed ambient fields. On the one hand, the reconnection heats the footpoints of the ambient fields with the left ones corresponding to the remote brightenings and the right ones mixed into the main flare brightenings. On the other hand, the reconnection also simultaneously transfers the materials from the rope-hosting filaments to newly formed ambient fields, and the movement of the filament materials to the remote brightenings gives rise to a red-shifted absorption in the H$\alpha$ profiles. Since the large-scale ambient loops are mainly connected to the sporadic negative polarities that are far away from the main active region and form ribbon-like patterns as shown in Figure \ref{fig:5}b, the footpoints of the large-scale ambient loops naturally appear as ribbon-like brightenings once the released energy from the interchange reconnection is transported there.

\begin{figure}[ht!]

{\centering\includegraphics[width=1.0\textwidth]{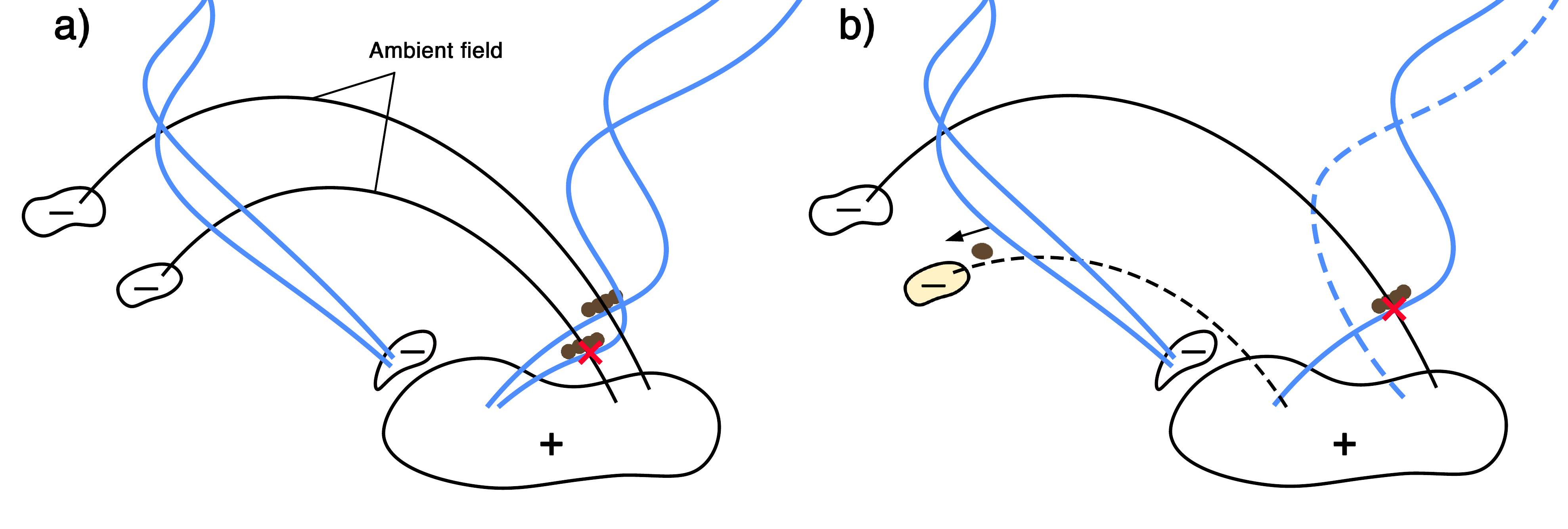}

}
\caption{Schematic interpretation for the formation of sequential remote brightenings. (a) The right leg of the erupting MFR (blue solid lines) reconnects with the closed ambient loops (black solid lines). The interchange reconnection location is denoted by the red cross. (b) After the interchange reconnection, a small amount of MFR flux (blue dashed line) and the newly formed ambient loops (black dashed line) change their footpoints. Filament materials (brown blob) from the MFR-hosting filament propagate along the newly formed loops to the remote brightening region (yellow). \label{fig:6}}
\end{figure}

\section{Summary and Discussions} \label{sec:cite}
In this study, we report imaging and spectroscopic observations of multiple ribbon-like remote brightenings pertinent to the main flare by CHASE/HIS and SDO. One significant property is that these remote brightenings occurred sequentially away from the main flare ribbons and were accompanied by downward-moving filament materials. These sequential remote brightenings are similar to sequential chromospheric brightenings first reported by \cite{Balasubramaniam2005} based on the common features, i.e., the sequential appearance at parasitic unipolar regions away from the main AR \citep{Balasubramaniam2005,Pevtsov2007}. However, some discrepancies are also found between them. In \cite{Balasubramaniam2005}, sequential chromospheric brightenings are composed of separated brightening points appearing at distinct locations and times. The propagation velocity of them is over 600 km s$^{-1}$. The remote brightenings observed here are more like sequential bright ribbons and have a slower propagation velocity of $\leq$ 300 km s$^{-1}$. The duration times of remote brightenings are also much longer than that (5.7 minutes) for sequential chromospheric brightenings \citep[e.g.,][]{Kirk2012b}. The sequential remote brightenings are also closely related to secondary flare ribbons, which are usually ribbon-like brightnings as extensions of the main flare ribbons \citep{Zhang2014}. We suspect that the secondary flare ribbons also originate from interchange reconnection but may correspond to the drifting footpoints of the erupting MFR as suggested in \citet{Lim2017} and \citet{Aulanier2019}. In contrast, the remote brightenings are argued to be the footpoints of newly-formed closed ambient loops.

The extrapolated 3D magnetic field configuration further justifies the interchange reconnection we conjecture. In Figure \ref{fig:6}, we interpret the process of interchange reconnection in more detail with a schematic drawing. As the MFR erupts outward, it encounters the ambient closed magnetic loops connecting the main AR positive polarity region and the sporadic negative polarities at the outskirts of the AR. The reconnection is supposed to occur between the MFR right leg and the large-scale ambient loops (Figure \ref{fig:6}a). When the interchange reconnection occurs, the ambient loops slightly change their footpoints close to the MFR footpoint. Meanwhile, the filament materials are also guided from the erupting MFR to the newly formed ambient loops, and the falling materials thus become visible in the H$\alpha$ red wing (Figure \ref{fig:6}b). Such a reconnection process is quite similar to the reconnection between the inclined arcade and the erupting MFR in \citet{Aulanier2019}. In \cite{Yan2020}, the similar reconnection process between the erupting filament and the overlying field was also used to explain the transfer and descent of the filament materials. Moreover, in the current event, there exists several groups of closed ambient loops whose remote footpoints are discretely distributed (Figure \ref{fig:5}a), which is the main reason of the remote brightenings appearing sequentially away from the main flare region.

We find that the bright loops (Figure \ref{fig:2}c-d) connecting the sequential remote brightenings to the main flare region have a quite long cooling time, which is estimated by the formula proposed by \citet{Cargill1995}:
\begin{equation}
\tau_{cool}\ \dot{=}\ 2.35 \times 10^{-2}L^{5/6}T_{e0}^{-1/6}n_{e0}^{-1/6} ,
\end{equation}
where $L$, $T_{e0}$ and $n_{e0}$ are the half-length, initial temperature and density of the loops, respectively. The initial temperature is regarded as $\sim 10^{7}$ K since the loops were first observed in the XRT images. The density is estimated to be about $ 9 \times 10^{9}$ cm$^{-3}$, as derived through $n_{e} = \sqrt{\frac{EM}{dl}}$ \citep{Cheng2012}, where EM is the total emission measure and $dl$ is the depth of the loops along the line-of-sight. The depth of the loops usually approximates their width, which is taken as 1.5 Mm here. Based on the extrapolated 3D magnetic field, the lengths of these loops range from 58 Mm to 121 Mm, the cooling times are thus calculated to be 1.3 - 2.5 hours, consistent with the observed cooling time.

Note that, the sequential remote brightenings as defined as sequential patches of EUV brightenings by \citet{Guo2015} were also suggested to be caused by magnetic reconnection between the expanding CME front and localized magnetic field. Nevertheless, for this particular event, such a possibility is excluded because both the CME front and associated Moreton wave were observed to travel across the remote brightening regions about 30 minutes earlier and that no any other waves are found to be temporally related to the remote brightenings.

Although we address that the sequential remote brightenings imply the interchange reconnection between the eruptive MFR and ambient field, which is even expected to widely exist in solar eruption events, such a reconnection process need more observational evidence. Further investigations with data-constrained MHD simulations are planned to reveal how an MFR (or a CME) interacts with the ambient fields and to determine the exact process of the interchange reconnection.

\begin{acknowledgments}
We would like to thank the referee for his/her constructive suggestions and Shihao Rao for valuable discussions. This work uses the data from CHASE mission supported by China National Space Administration. B.T.W., X.C., C.L., J.C. and M.D.D. are funded by NSFC grant 12127901 and by National Key R\&D Program of China under grant 2021YFA1600504.
\end{acknowledgments}

\bibliography{Total}{}
\bibliographystyle{aasjournal}

\end{document}